\documentstyle[epsfig,psfig]{texas}

\def\msun{M_\odot}
\def\lappreq{{{<}\atop{\sim}}}
\def\gappreq{{{>}\atop{\sim}}}

\begin{document}
\title{STOCHASTIC BACKGROUNDS OF GRAVITATIONAL WAVES 
FROM COSMOLOGICAL POPULATIONS OF ASTROPHYSICAL SOURCES}

\author{Raffaella Schneider$^{1}$, Valeria Ferrari$^{1}$, 
Sabino Matarrese$^{2}$}
\address{(1) Dipartimento di Fisica, Universit\'a degli Studi di Roma, 
``La Sapienza'' and Sezione INFN Roma1 \\
Piazzale Aldo Moro 5, 00185 Roma, Italy\\}
\address{(2) Dipartimento di Fisica, Universit\'a degli Studi di Padova and
Sezione INFN Padova\\
Via Marzolo 8, 35131 Padova, Italy\\}
{\rm Email: schneider@roma1.infn.it, valeria@roma1.infn.it, 
matarrese@pd.infn.it}

\begin{abstract}
Astrophysical sources of 
gravitational radiation are likely to have been formed since the beginning
of star formation.\\ 
Realistic source rates of formation throughout the Universe have been 
estimated from an observation-based determination of the star formation rate
density evolution.\\ 
Both the radiation emitted during the collapse to black holes and the 
spin-down
radiation, induced by the r-mode instability, emitted by hot, 
young rapidly rotating neutron stars have been
considered.\\
We have investigated the overall signal produced by the ensemble of sources
exploring the parameter space and discussing its possible detectability.
\end{abstract}

\section{Introduction}
Stochastic backgrounds of gravitational waves are interesting sources
for the interferometric detectors that will soon start to operate.\\
Their production is a robust prediction of any model which attemps
to describe the evolution of the Universe at primordial epochs.\\
However, bursts of gravitational radiation emitted by a large number of
unresolved and uncorrelated astrophysical sources generate a stochastic
background at more recent epochs, immediately following the onset of galaxy
formation.\\
Thus, astrophysical backgrounds might overwhelm the primordial ones
and their investigation provides important constraints on the
detectability of signals coming from the very early Universe.\\
The main characteristics of the gravitational backgrounds produced by 
cosmological populations of astrophysical sources depend both on the
emission properties of each single source and on the source rate evolution
with redshift.\\
Extra-galactic backgrounds are proved to be mainly contributed by sources
at redshifts $z \leq 1-2$ and their formation rate can not be simply 
extrapolated from its local value but must account for the evolution of the 
overall galactic population \cite{FMS99a}, \cite{FMS99b}, \cite{KP98}.   
The model we have adopted for the redshift evolution of the source rate of
formation is described in Section 1 and it is based on the star formation 
history derived by UV-optical observations of star forming galaxies out to 
redshifts of $\sim 4-5$ (see e.g. \cite{M99}, \cite{LBG98}).\\
The gravitational wave sources for which the extra-galactic background
contributions have been investigated so far are white dwarfs binary systems   
during the early in-spiral phase \cite{KP98} and core-collapse SNae. 
In particular, we have considered the
gravitational waves emitted during the core-collapse to a black hole 
\cite{FMS99a} and the gravitational radiation emitted by newly formed, 
rapidly rotating, hot neutron stars with an instability in their r-modes
\cite{FMS99b}. The first choice was motivated by the results of numerical
simulations of core-collapses: unlike the case of a core-collapse to a neutron
star, the gravitational wave emission spectrum produced during a core-collapse
to a black hole is rather generic, in the sense that it is sufficiently 
independent of the initial conditions and of the equation of state of the 
collapsing star (see, for a recent review, \cite{FP98}). 
The second kind of sources were considered because of their
high efficiency in producing gravitational signals: though preliminary, 
the investigations of the r-modes instabilities in highly rotating
young neutron stars have proved that a considerable fraction 
of the star initial
rotational energy is converted in gravitational waves, making the process
very interesting for gravitational wave detection \cite{A98},
\cite{FM98}, \cite{LOM98}, \cite{O98}, \cite{AKS98}.\\
A brief description of the characteristics of the source emission spectra is 
given in Section 2. Finally, in Section 3 we derive the spectra of the 
corresponding backgrounds, explore the 
parameter space and discuss their detectability.

\section{The source formation rate}

\begin{figure}
\centering
\psfig{file=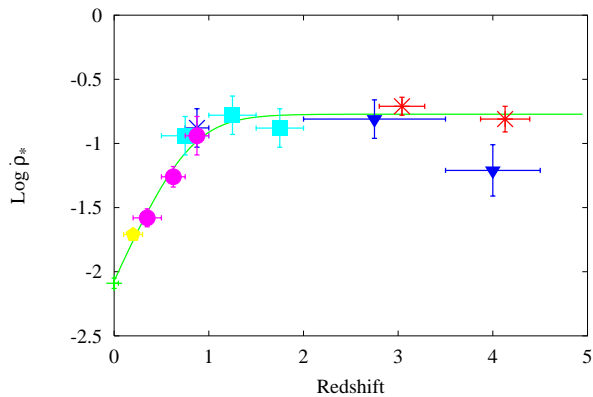,angle=270,width=8cm}
\caption{The Log of the star formation rate density in units of 
$\msun \mbox{yr}^{-1} \mbox{Mpc}^{-3}$ as a function of redshift for a 
cosmological background model with  $\Omega_{M}=1$, $\Omega_{\Lambda}=0$,
$H_0=50 \,\mbox{km}\mbox{s}^{-1} \mbox{Mpc}^{-1}$ and a 
Salpeter IMF (see text). The data points 
correspond to \cite{G95} (cross), \cite{T98} (filled pentagon), \cite{CFRS}
(filled circles), \cite{C97} (filled squares), HDF (\cite{M96}, filled 
triangles), \cite{LBG96}, \cite{LBG98} (asterisks).} 
\label{sfr}
\end{figure}

\begin{figure}
\centering
\psfig{file=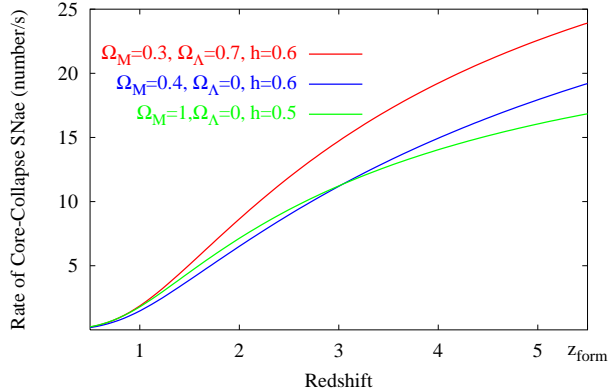,angle=270,width=8cm}
\caption{The rate of core-collapse supernovae (progenitor masses in the range
$8 \msun \lappreq M \lappreq 125 \msun$) for three different cosmological
background models. When varying the cosmological parameters, both 
the comoving volume element and the star formation rate density are
properly modified (see \cite{FMS99a})}.  
\label{rate}
\end{figure}

In the last few years, the extraordinary advances attained in observational
cosmology have led to the possibility of identifying actively 
star forming galaxies at increasing cosmological look-back times 
(see e.g. \cite{E97}).
Using the rest-frame UV-optical luminosity as an indicator of the star
formation rate and integrating on the overall galaxy population,
the data obtained with the {\it Hubble Space Telscope} (HST \cite{M96},
\cite{C97}) Keck and
other large telescopes \cite{LBG96}, \cite{LBG98} together with the 
completion of several large redshift surveys \cite{G95}, \cite{T98}, 
\cite{CFRS} have enabled, for the first time, to derive coherent models for
the star formation rate evolution throughout the Universe. \\
A collection of some of the data obtained at different redshifts together 
with a proposed fit is shown in  Figure~\ref{sfr}. 
Because dust extinction can lead to
an underestimate of the real UV-optical emission and, ultimately, of the 
real star formation activity, the data shown in Fig.~\ref{sfr} 
have been corrected
upwards according to the factors implied by the Calzetti
dust extinction law (see \cite{DC}, \cite{LBG98}). 
Although the strong luminosity 
evolution
observed between redshift 0 and 1-2 is believed to be quite firmly established,
the amount of dust correction to be applied at intermediate redshift (thus
the amplitude of the curve at $z \sim 1-2$) as well as the behaviour of the
star formation rate at high redshift is still relatively uncertain.
In particular, the decline of the star formation rate density implied by
the $<z> \sim 4$ point of the {\it Hubble Deep Field} 
(HDF, see Fig.~\ref{sfr}) 
is now
contradicted by the star formation rate density derived from a new sample of 
Lyman break galaxies with $<z>=4.13$ \cite{LBG98} 
which, instead, seems to indicate that the star
formation rate density remains substantially constant at $z>1-2$.
It has been suggested that this discrepancy might be caused by problems
of sample variance in the HDF point at $<z>=4$ \cite{LBG98}. \\
Thus, we have up-dated the star formation rate model that we have previously
considered in the analysis  even though the
gravitational wave backgrounds are almost insensitive to the behaviour
of the star formation rate at $z>1-2$ because the contributions of very
distant sources is very weak \cite{FMS99a}, \cite{FMS99b}. Conversely,
if a larger dust correction factor should be applied at intermediate
redshifts, this would result in a similar amplification of the 
gravitational background spectra.\\
From the star formation history plotted in Figure~\ref{sfr}, 
it is possible to infer the formation rate 
(number of objects formed per unit time) of a particular population of 
gravitational wave sources (remnants) 
by integrating
the star formation rate density over the comoving volume element out to 
redshift $z$ and considering only those progenitors with masses 
falling in the correct dynamical range for the remnant to form, i.e.,
\begin{equation}
R(z)=\int_{0}^{z} dz' \, \frac{\dot{\rho}_{\star}(z')}{1+z'} \, \frac{dV}{dz}
\, \int_{\Delta M} \!\! dM' \, \Phi(M'),
\end{equation}
where the factor $(1+z)^{-1}$ takes into account the dilution due to cosmic
expansion and $\Phi(M)$ is the initial mass function (IMF) chosen to be 
of Salpeter type, $\Phi(M) \propto M^{-(1+x)}$ with $x=1.7$.\\
Stellar evolution models have shown that single stars with masses 
$\geq 8 \msun$ pass through all phases of nuclear burning and end up as
core-collapse supernovae leading to a neutron star or a black hole remnant.
While there seems to be a general agreement that progenitors with
masses in the range $8 \msun \lappreq M \lappreq 20 \msun$ leave neutron star
remnants, the value of the minimum progenitor mass which leads to a black hole
remnant is still uncertain, mainly because of the unknown amount of 
fall back of material during the supernova explosion 
\cite{WW95}, \cite{WT96}. In our analysis, a reference interval of
$25 \msun \lappreq M \lappreq 125 \msun$ was considered but we have
also investigated the effects of choosing a lower limit of $20 \msun$ and 
$30 \msun$ as well as an upper limit of $60 \msun$ \cite{FMS99a}.\\
The rate of core-collapse SNae predicted for three cosmological background
models is shown in Figure~\ref{rate} as a function of redshift.
The main
difference between the three cosmologies is introduced by the 
geometrical effect of the comoving volume and is significant at
$z \gappreq 1-2$. This implies that the gravitational backgrounds, which
are mainly contributed by sources at $z \lappreq 1-2$,  
are almost insensitive to the cosmological parameters.\\
The total black hole formation rate $R_{BH}$  
and neutron star formation rate $R_{NS}$ predicted by our model are,
\begin{equation}
R_{BH}= 3.3 - 4.7 \mbox{s}^{-1} \qquad R_{NS}=13.6 - 19.3 \mbox{s}^{-1}
\end{equation}
depending on the cosmological background model considered. The value
predicted by our model for the local core-collapse SNa rate is in good
agreement with the available observations \cite{Capp97}. \\

\section{The single source emission spectra}

\begin{figure}
\centering
\psfig{file=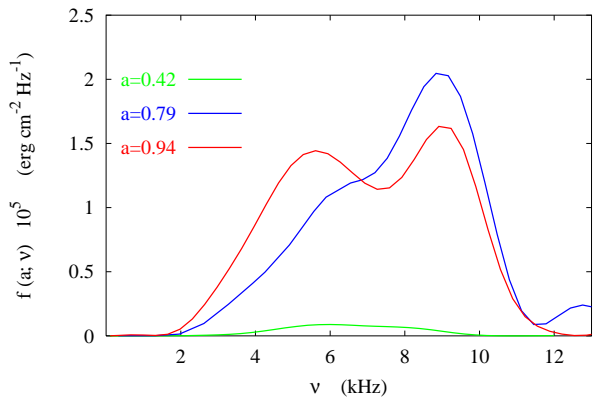,angle=270,width=8cm}
\caption{The average energy flux emitted during the axisymmetric collapse
of a rotating, polytropic star to a black hole of $M_{core}=1.5 \, \msun$ at
a distance of $15$ \, Mpc. The three curves correspond to assigned values of 
the rotational parameter}.  
\label{bhsingle}
\end{figure}

The emission spectrum that we have adopted as our model
for the gravitational wave radiated from a core which is 
collapsing to a black hole was that obtained from a 
fully non-linear numerical simulation of Einstein+hydrodynamic
equations of an axisymmetric core collapse \cite{SP85}, \cite{SP86}.\\
The main properties of the spectrum are shown in Fig.~\ref{bhsingle}
for the collapse of a $1.5 \msun$ naked core to a 
black hole at a distance of $15$ Mpc 
and for three assigned values of the angular momentum. \\	
The relevant quantity is the rotational parameter $a=J/(GM_{core}^2/c)$.
In fact, there is a maximum in the emission at a frequency which
 depends on the value of $a$ and whose amplitude, for values
of $a$ in the range $0.2 < a < 0.8$, scales as $a^4$. This peak 
is located at a
frequency which is very close to the frequency of the
lowest $m=0$ quasi-normal mode.
This means that a substantial fraction of the  energy
will be emitted after the black hole has formed: 
it will oscillate in its quasi-normal modes 
until its residual mechanical energy is  radiated away in gravitational
waves.\\
For high values of the rotational parameter, the geometry of the collapse
is different as the star becomes flattened into the equatorial plane
and then bounces vertically, but still continues to collapse
inward until the black hole is formed. In this case,
a low frequency component appears, with an  
amplitude which may become comparable to that of
the peak corresponding to the quasi-normal modes.\\
In general, the efficiency of this  
axisymmetric core-collapse to a black hole
is $\Delta E_{GW}/M_{core} c^2 \leq 7 \times 10^{-4}$. 
It should be remembered that less symmetric configurations may result in a 
more efficient production of gravitational waves.\\

A number of investigations of relativistic rotating stars
has recently led to the discovery of a new class of instability modes, called
the r-modes \cite{A98}, \cite{FM98}, \cite{LOM98}, \cite{O98}, \cite{AKS98}.\\
These modes are characterized by having the Coriolis force as the restoring
force and thus they are relevant only for rotating stars. Even though the
analyses carried out so far are still preliminary and are based on several
approximations, these modes, whose instability is driven by gravitational 
radiation, appear to efficiently radiate in gravitational waves a large
part of the initial rotational energy in a relatively small time interval.\\
A preliminary estimate of the corresponding  emission spectrum was 
recently obtained in \cite{O98} for a polytropic neutron star model with
a $1.4 \msun$ mass and a radius of $12.53$ km. We have adopted their
proposed spectrum as our model for the single source emission 
in order to estimate the gravitational background produced by
young, hot, rapidly rotating neutron stars through the r-mode instability 
\cite{FMS99b}. \\
The evolution of the angular momentum of the star is determined
by the emission of gravitational waves, which couple to the r-modes
through the current multipoles, primarily that with $l=m=2$.
For this mode, the frequency of the
emitted gravitational radiation is $\nu=(2/3\pi)\,\Omega$.
The star is assumed to be initially rotating at its maximum spin rate, i.e.,
at its Keplerian value $\Omega_K$, which corresponds to a gravitational wave
frequency of $\sim 1400$ km, for the star model considered \cite{O98}.
The evolution of $\Omega$ during the phase
in which the amplitude of the mode is small can be determined 
from the standard multipole expression for angular momentum loss, and
from the energy loss due to the gravitational emission and to the  dissipative
effects induced by the bulk and shear viscosity.
In this phase, $\Omega$ is nearly constant and
the instability grows exponentially. After a short time, the
amplitude of the mode becomes close to unity and non-linear 
effects saturate and halt further growth of the mode. This phase
lasts for approximately 1 yr, during which the star loses
angular momentum radiating approximately $2/3$ of its
initial rotational energy in gravitational waves,
up to a point where the angular velocity reaches a critical value, $\Omega_c$.
This value can be determined by solving the equation 
$1/\tau(\Omega_c)=0$, where
$\tau$ is the total dissipation time-scale which can be decomposed as a sum 
of the damping times associated to the gravitational emission,  to the
shear and to the bulk viscosity. 
$\tau(\Omega_c)$ is clearly  a function of the temperature of the star,
and it has been shown that the r-mode instability operates only in hot
neutron stars ($10^{10}\gappreq T \gappreq 10^9$ K) 
\cite{O98}. Above $10^{19}$ bulk viscosity kills the r-mode instability 
whereas below $10^9$ K superfluidity
and other non-perfect fluid effects become important and the
damping  due to  viscosity dominates with respect to the destabilizing
effect of the gravitational radiation. For the star model considered,
$\Omega_c \approx 566$ Hz,  
which corresponds to a final spin period of $\approx 11$ ms and to
$\nu_{min} \approx 120$ Hz. Below this critical value,
viscous forces and gravitational radiation damp out the energy left
in the mode, and the star slowly reaches its final equilibrium configuration.\\
The qualitative picture that arises from this simple model is believed to
be sufficiently reliable, even though various uncertainties and approximations
might affect the quantitative results for the initial rotation of the star
after collapse, for the spin-down time-scales 
as well as for the final rotation period \cite{AKS98}. However, in this 
framework the expression of the  energy spectrum  can be approximated as 
follows,
\begin{equation}
\frac{dE_{GW}}{d\nu} \approx \frac{4}{3} \, E_K\, \frac{\nu}{\nu_{max}^2}
\qquad  \mbox{for} \qquad \nu_{min} \leq \nu \leq \nu_{max}
\end{equation}
where $E_K$ indicates the initial rotational energy \cite{O98}.\\
Thus, the mean flux emitted by this source can be written as,
\begin{equation}
f(\nu)=\frac{1}{4\pi d^2} \left(\frac{d E_{GW}}{d\nu}\right).
\end{equation}

\section{The stochastic backgrounds}

\begin{figure}
\centering
\psfig{file=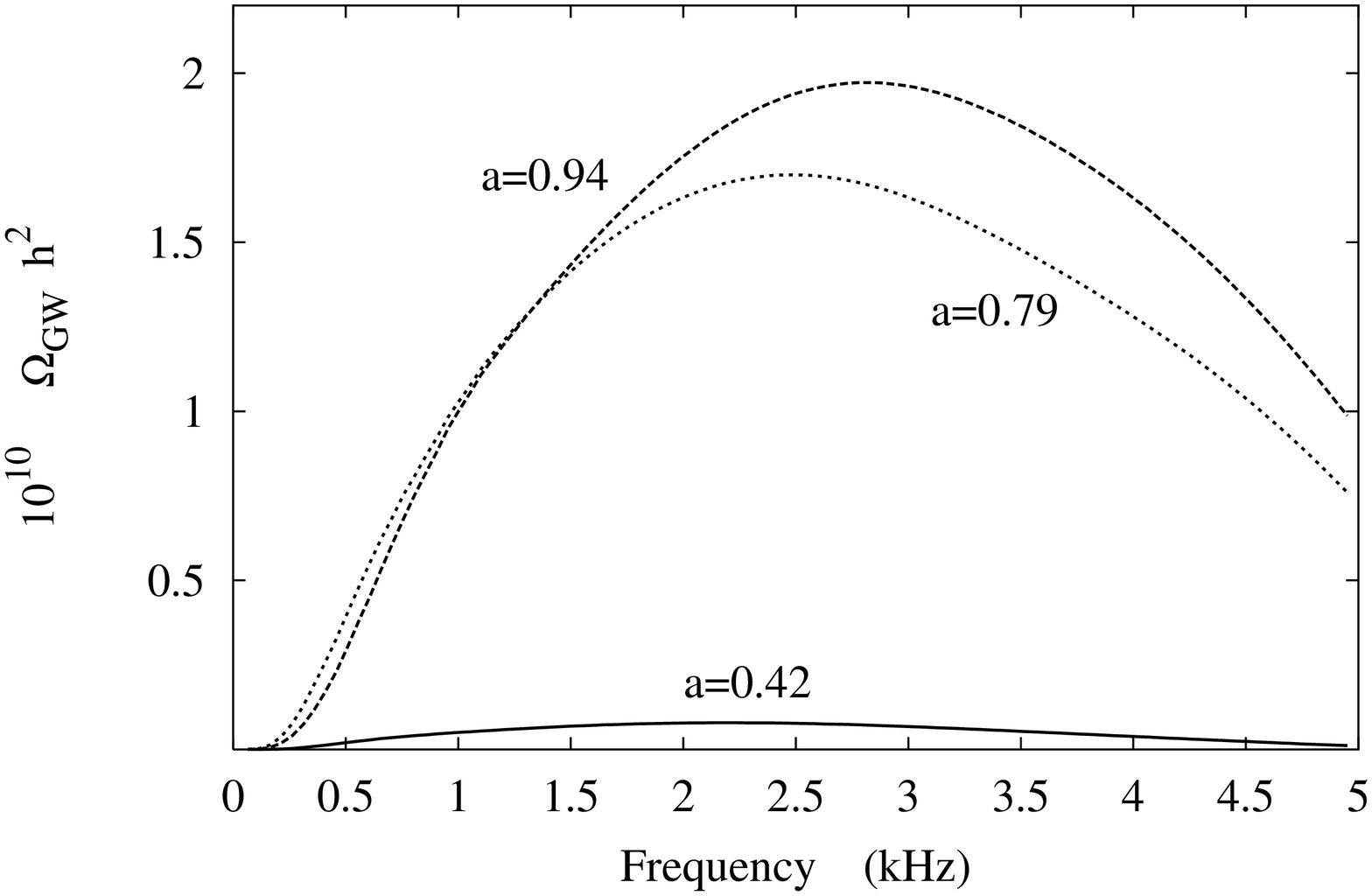,angle=270,width=6cm}
\psfig{file=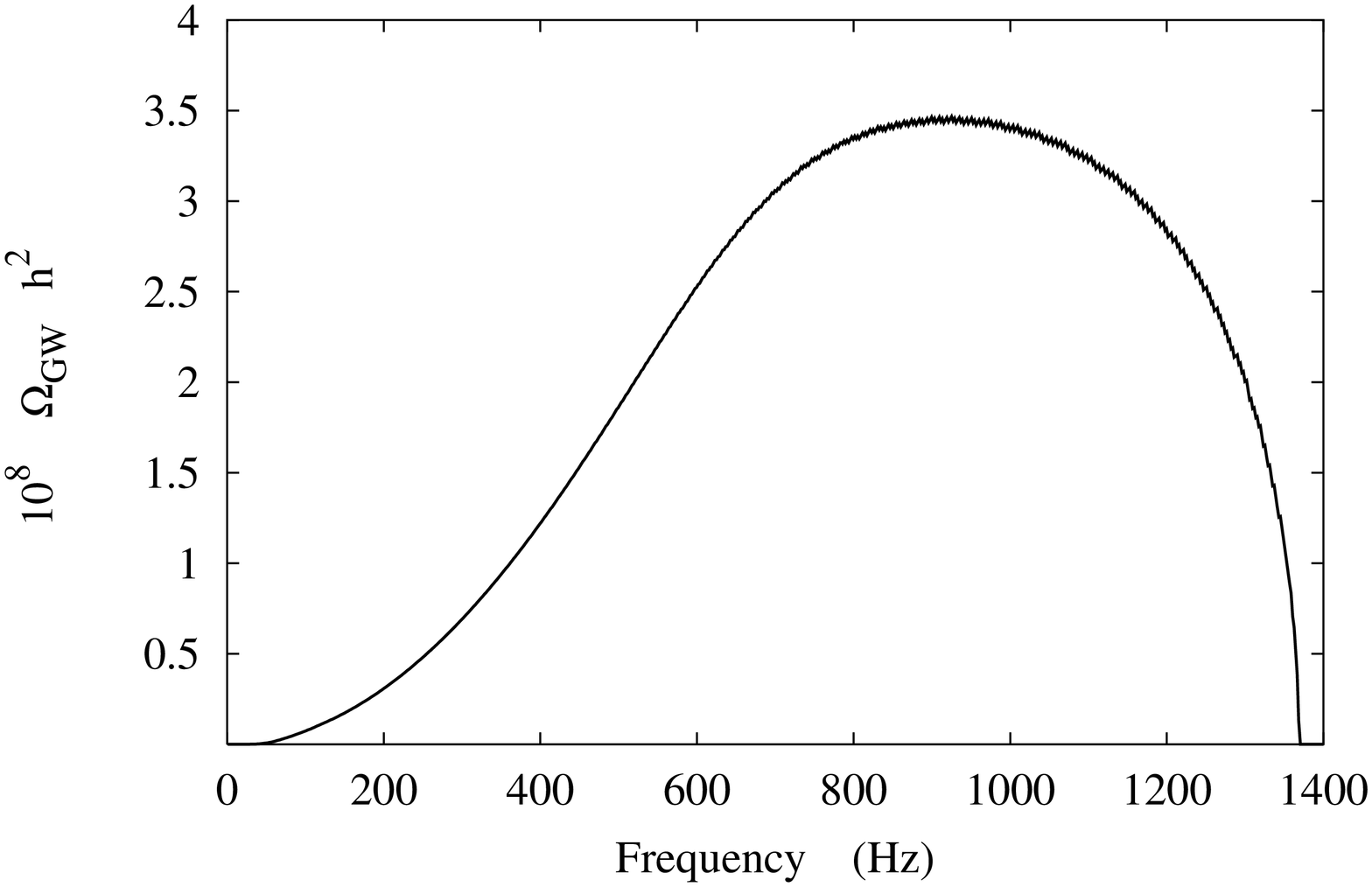,angle=270,width=6cm}
\caption{The functions $h^2\,\Omega_{GW}$ vs the 
observational frequency corresponding to the background produced
by the radiation emitted by an ensemble of axisymmetric black hole
collapses (left panel) and by the spin-down radiation emmited
by young rapidly rotating neutron stars (right panel). 
The three curves in the left panel correspond
to assigned values of the rotational parameter. A flat cosmological
background model with zero cosmological constant and $h= H_0/100 = 0.5$
is assumed (see text).} 
\label{omega}
\end{figure}

In order to evaluate the spectral energy density, 
$ dE/dt dS d\nu$, of the stochastic backgrounds
produced by the radiation emitted during an axisymmetric black hole collapse
and by the spin-down radiation from newly born
neutron stars, we need to convolve the differential
rate of sources, $dR(z)$, with the flux emitted by a single source at 
redshift $z$ as it would be observed today (see \cite{FMS99a}, \cite{FMS99b}).
This means that we account for the luminosity distance damping on the flux
emitted by a single source and we redshift the emission frequencies.\\
The corresponding values of the closure energy densities of gravitational waves
can be obtained as follows, 
\begin{equation}
\Omega_{GW}(\nu_{obs})= \frac{\nu_{obs}}{c^3 \rho_{\mbox{cr}}}\, \frac{dE}{dt dS
d\nu},
\end{equation}
where $\rho_{\mbox{cr}}= 3H_0^2\,/\,8\pi ~G$ and are shown in Fig.~\ref{omega}. 
These Figures have been obtained for a flat cosmological background model
with zero cosmological constant and with a Hubble constant of $H_0=
h \, 100 = 50 \mbox{km}/\mbox{s}^{-1} \mbox{Mpc}^{-1}$.\\
As previously mentioned, the effect of a varying cosmological background 
is negligible on the final properties of the stochastic backgrounds. 
In fact, the amplification
of the rate at high redshifts shown in Fig.~\ref{rate} 
for an open model and a model with a cosmological constant
is mostly suppressed by
the inverse squared luminosity distance dependence of the single
source spectrum for the same models. \\
The closure density of the black hole collapse background is shown in the
left panel of Fig.~\ref{omega} 
for three values of the rotational parameter. Since we do not
know the distribution of angular momenta, for each curve all the sources
of the ensemble were assumed to have the same value of $a$. Depending on this
value, the closure density has a maximum amplitude in the range 
$\sim 10^{-9}-10^{-10}$ at frequencies between $\sim 2-3$ kHz. 
Even though the final properties of the background depend on the
model that we have assumed as being representative of the process of 
gravitational collapse to a black hole, the relevant features of the energy 
spectrum  we use to model each single event are likely to reasonably 
represent a generic situation, (see the discussion in \cite{FMS99a}).
As for the dependence on the formation rate of black holes, 
the uncertainties which affect the evolution of the star formation rate at 
high redshifts are completely irrelevant whereas variations induced by
different lower and upper mass cut-offs of the progenitor mass range are
limited to a factor $\lappreq 2$ \cite{FMS99a}.\\
As shown in the right panel of Fig.~\ref{omega}, 
the closure density for the neutron star
background has a larger amplitude than the previous case and the main part of 
the signal is concentrated at lower frequencies. In fact, it is
characterized by a wide maximum, ranging from $\sim 0.7 -1$ kHz, with an
amplitude of a few $10^{-8}$. Allowing for variations in $\nu_{min}$ and 
$\nu_{max}$ does not substantially alter the main features of the background
although some quantitative differences appear both in the small and large
frequency part of the signal (see Fig.s 6 and 8 in \cite{FMS99b}).\\
The neutron star background allows a clear inspection
of the impact of the star formation rate evolution on its final properties.
In fact, in this case all the sources have been assumed to have 
the same mass and thus, elements
of the ensemble at the same redshift have exactly the same emission
properties. Therefore, it is easier to distinguish the effects of the source 
rate evolution from that of the spectrum of each single event. 
\begin{figure}
\centering
\psfig{file=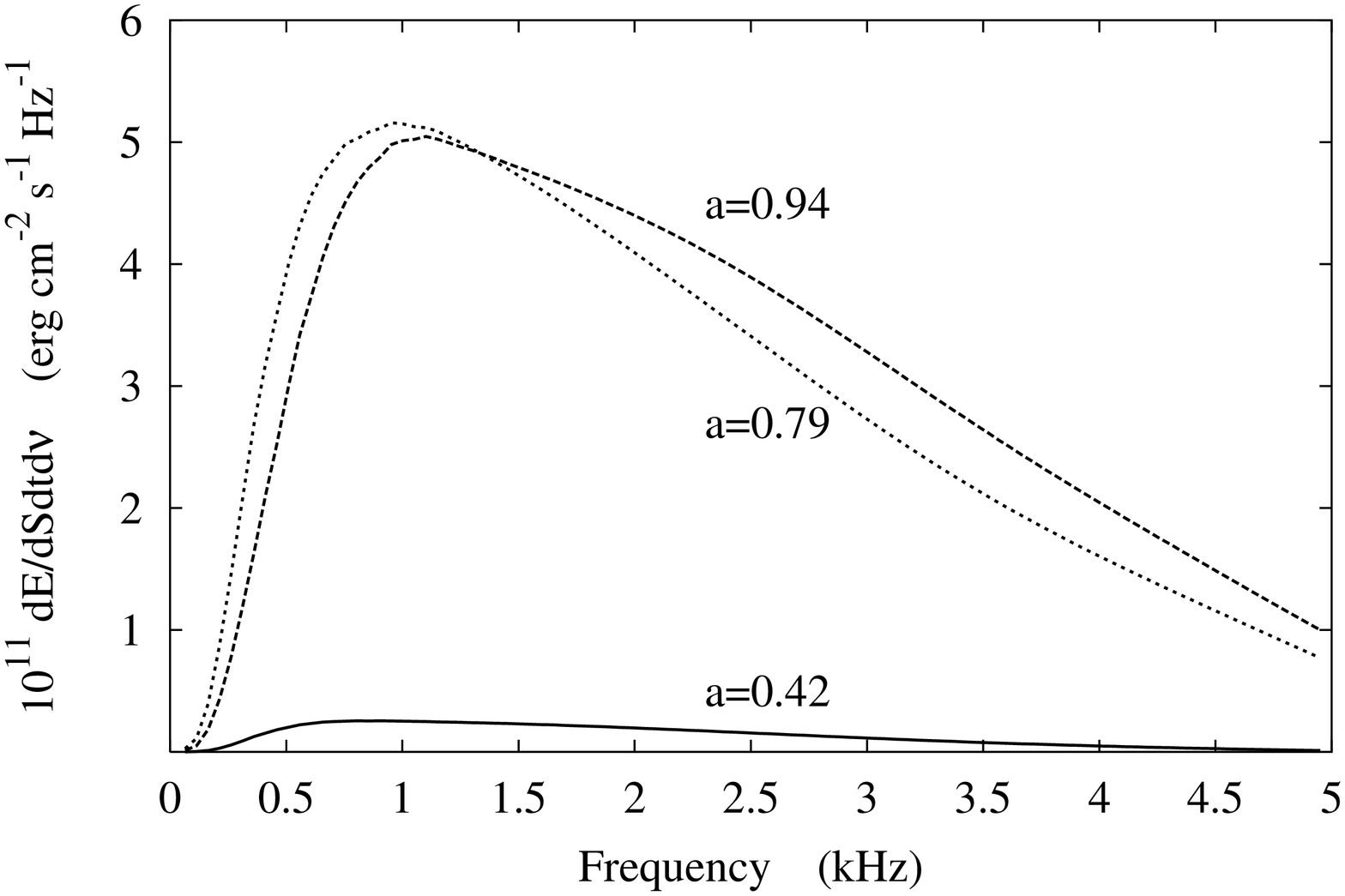,angle=270,width=6cm}
\psfig{file=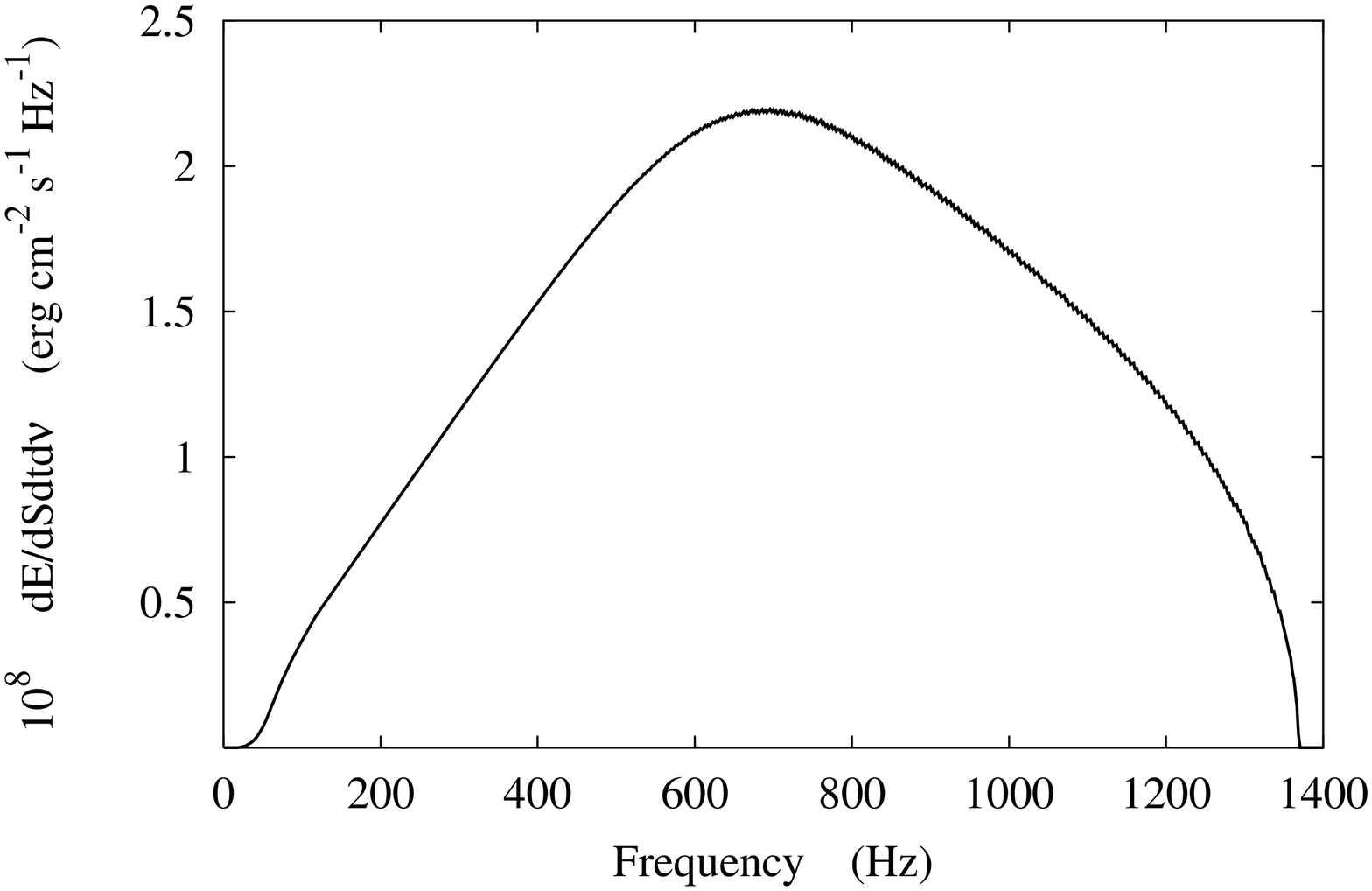,angle=270,width=6cm}
\caption{The spectra vs the 
observational frequency corresponding to the background produced
by the radiation emitted by an ensemble of axisymmetric black hole 
collapses (left panel) and by the spin-down 
radiation emitted by an ensemble of rapidly rotating neutron star (right
panel). The three curves on the left panel correspond
to assigned values of the rotational parameter. A flat cosmological
background model with zero cosmological constant and $h= H_0/100 = 0.5$ 
is assumed (see text).} 
\label{spectra}
\end{figure}
The right panel of Fig.~\ref{spectra} shows the spectrum of 
the neutron star background.
The maximum amplitude occurs around $\sim 700$ Hz. This means that the
most significant contribution to the background signal comes from neutron
stars at their maximum spin rate ($\sim 1400$ Hz, for our model) which
are formed at redshifts $z \sim 1-2$ where the star formation rate
reaches its maximum value before entering its high redshift plateau.
Similarly, if one takes into account that the mean value of the core mass which
collapses is around $\approx 4-5 \msun$, the corresponding maximum 
in the contribution of a mean single source occurs at rest-frame 
frequencies in the 
range $2-3$ kHz.
From the left panel of Fig.~\ref{spectra} it is possible to see that the 
maximum amplitude
in the black hole background spectra corresponds to frequencies $\approx 1-2$ 
kHz, depending on the value of the rotational parameter. 
Thus, the relevant contribution to the final black hole background signal
comes from those sources which are formed around $z \sim 1-2$.\\
Moreover, it is important to note that for sources, such as the one we
have described, which emit gravitational waves at rest-frame frequencies 
$\nu \gappreq 100$ Hz, at frequencies $1-100$ Hz, where cross-
correlation between terrestrial interferometers can be accomplished,       
the stochastic background signal is entirely produced at $0<z<1-2$. \\
We can conclude that a reliable estimate of astrophysical backgrounds
can not set the important effect of the star formation rate evolution aside.\\
Finally, it is possible to show that the first generation of interferometers
will not reach the sensitivities required to observe these backgrounds.
In fact, the relevant part of the signal is at relatively high 
frequencies where, at their actual sites, the interferometers that will soon 
start to operate can not be cross-correlated.\\
For the first generation of interferometers, the best signal-to-noise ratio 
is obtained by cross-correlating VIRGO and GEO600 optimally oriented. 
Assuming one year of integration, $\mbox{S}/\mbox{N} \sim 2 \times 10^{-3}$.
For the same integration time, two LIGO interferometers with advanced 
sensitivities give $\mbox{S}/\mbox{N} \sim 1.23$ at their actual sites and
$\mbox{S}/\mbox{N} \sim 15$ if they were at a distance of $\sim 300$ km.\\
Though signal-to-noise ratios calculated for intereferometer-bar pairs, such as
VIRGO-NAUTILUS or GEO600-NAUTILUS are still very low, two hollow spheres
with $\sqrt{S_n(200 \, \mbox{Hz})} \sim 10^{-24}$ placed at the same site
would reach, in one year of integration, a signal-to-noise ratio 
$\mbox{S}/\mbox{N} \sim 1$ \cite{TVG}.
So far, the stochastic backgrounds we have described were 
considered to be continuous. This is always the case for the background
produced by the spin-down radiation emitted by rapidly rotating neutron stars,
as the signal from each single source is emitted in a relatively long
time interval, of the order of 1 yr (see Section 2). Thus, these signals
can superimpose and do form a continuous background \cite{FMS99a}. 
Conversely, the background produced by core collapses to black holes 
has a shot noise character. 
In fact, the typical duration of the gravitational 
signal emitted by each source is much shorter than the previous case, of
the order of a ms. Thus, the contributions from the elements of the ensemble
do not superimpose but rather generate a shot-noise background, characterized
by a succession of isolated bursts with a mean separation of the order of
$0.1$ seconds,
much longer than the typical duration of each burst \cite{FMS99a}.
The peculiar statistical character of this background might be
exploited in order to design a specific algorithm which may help its detection.

\end{document}